\documentclass[10pt]{article}

\usepackage{palatino}

\usepackage{url}
\usepackage{xcolor}
\usepackage{color}
\usepackage{graphicx}
\usepackage{amssymb}
\usepackage{amsmath,amsthm,paralist}
\usepackage{amsbsy}
\usepackage{cite}
\usepackage{algorithm,algpseudocode}
\usepackage{comment}

\def\real    { \mathbb{R} }
\def\complex{\mathbb{C}}

\theoremstyle{remark}

\newcommand{\rank}{\text{rank}}
\newcommand{\trace}{\text{trace}}
\newcommand{\inner}[1]{\left<#1\right>}
\DeclareMathOperator*{\argmin}{\text{arg~min}}

\DeclareMathOperator*{\minimize}{\text{minimize}}

\def\R{R}
\def\obs{L}
\def\M{M}
\def\N{N}
\def\bX{{\mathbf X}}
\def\target{\bX_0}
\def\bA{{\mathbf A}}
\def\cA{{\mathcal A}}
\def\by{{\mathbf y}}

\def\cP{{\mathcal P}}

\def\bu{{\mathbf u}}
\def\bv{{\mathbf v}}

\def\bz{{\mathbf z}}

\def \Real {\mathbb{R}}

\graphicspath{{./Figs/}}

\renewcommand{\Re}[1]{\operatorname{Re}\left\{#1\right\}}
\renewcommand{\Im}[1]{\operatorname{Im}\left\{#1\right\}}
\newcommand{\conj}[1]{\mkern 1.5mu\overline{\mkern-1.5mu#1\mkern-1.5mu}\mkern 1.5mu}

\renewcommand{\P}[1]{\operatorname{P}\left(#1\right)}
\newcommand{\E}{\operatorname{E}}

\newcommand{\e}{\mathrm{e}}
\renewcommand{\j}{\mathrm{j}}

\newcommand{\vct}[1]{\boldsymbol{#1}}
\newcommand{\mtx}[1]{\boldsymbol{#1}}
\renewcommand{\H}{\mathrm{H}}
\newcommand{\T}{\mathrm{T}}

\newcommand{\Null}{\operatorname{Null}}

\newcommand{\set}[1]{\mathcal{#1}}

\newcommand{\va}{\vct{a}}

\newcommand{\ve}{\vct{e}}
\newcommand{\vf}{\vct{f}}

\newcommand{\vh}{\vct{h}}

\newcommand{\vs}{\vct{s}}

\newcommand{\vu}{\vct{u}}
\newcommand{\vv}{\vct{v}}
\newcommand{\vw}{\vct{w}}

\newcommand{\vy}{\vct{y}}

\newcommand{\vlambda}{\vct{\lambda}}

\newcommand{\mA}{\mtx{A}}
\newcommand{\mB}{\mtx{B}}

\newcommand{\mE}{\mtx{E}}

\newcommand{\mG}{\mtx{G}}
\newcommand{\mH}{\mtx{H}}

\newcommand{\mL}{\mtx{L}}

\newcommand{\mR}{\mtx{R}}

\newcommand{\mU}{\mtx{U}}
\newcommand{\mV}{\mtx{V}}

\newcommand{\mX}{\mtx{X}}
\newcommand{\mY}{\mtx{Y}}
\newcommand{\mZ}{\mtx{Z}}

\newcommand{\mSigma}{\mtx{\Sigma}}

\newcommand{\mzero}{{\bf 0}}

\newcommand{\setB}{\set{B}}
\newcommand{\setC}{\set{C}}

\newcommand{\setQ}{\set{Q}}
\newcommand{\setR}{\set{R}}
\newcommand{\setS}{\set{S}}
\newcommand{\setT}{\set{T}}

\newboolean{twoColVersion}
\setboolean{twoColVersion}{false}
\newcommand{\twoCol}[2]{\ifthenelse{\boolean{twoColVersion}} {#1} {#2} }

\begin{document}

\title{An overview of low-rank matrix recovery from incomplete observations}

\vspace{2mm}
\author{Mark A.\ Davenport and Justin Romberg\footnote{M.\ D.\ and J.\ R.\ are with the School of Electrical and Computer Engineering, Georgia Institute of Technology, Atlanta, GA 30332 USA (e-mail: \{mdav,jrom\}@gatech.edu).
M.\ D.\ is supported by NRL grant N00173-14-2-C001, AFOSR grant FA9550-14-1-0342, and NSF grants CCF-1409406, CCF-1350616, and CMMI-1537261.  J.\ R.\ is supported by ONR grant N00014-11-1-0459, NSF grant CCF-1422540, and a grant from the Packard Foundation.}}
\date{January 24, 2016}

\maketitle

\begin{abstract}
Low-rank matrices play a fundamental role in modeling and computational methods for signal processing and machine learning.  In many applications where low-rank matrices arise, these matrices cannot be fully sampled or directly observed, and one encounters the problem of recovering the matrix given only incomplete and indirect observations.  This paper provides an overview of modern techniques for exploiting low-rank structure to perform matrix recovery in these settings, providing a survey of recent advances in this rapidly-developing field.  Specific attention is paid to the algorithms most commonly used in practice, the existing theoretical guarantees for these algorithms, and representative practical applications of these techniques.
\end{abstract}

%----------------------------------------------------------------------------------
\section{Introduction}

Low-rank matrices arise in an incredibly wide range of settings throughout science and applied mathematics.  To name just a few examples, we commonly encounter low-rank matrices in contexts as varied as:
\begin{itemize}
\item {\bf ensembles of signals}: the output of a sensor array or network, a collection of video frames, or a sequence of segments of a longer signal can often be highly correlated and represented using a low-rank matrix~\cite{DavieE_Rank,ahmed15co};
\item {\bf system identification}: low-rank (Hankel) matrices correspond to low-order linear, time-invariant systems~\cite{LiuV_Interior};
\item {\bf adjacency matrices}: the connectivity structure of many graphs, such as those that arise in manifold learning and social networks, is often low rank~\cite{LiniaLR_Geometry};
\item {\bf distance matrices}: in many data embedding problems --- such as those that arise in the context of multidimensional scaling~\cite{BorgG_Modern}, sensor localization~\cite{SoY_Theory,BiswaLWY_Semidefinite}, nuclear magnetic resonance spectroscopy~\cite{Singe_Remark,SingeC_Uniqueness}, and others --- the matrix of pairwise distances will typically have a rank dependent on the (low) dimension of the space in which the data lies;
\item {\bf item response data}: low-rank models are frequently used in analyzing data sets containing the responses of various individuals to a range of items, such as survey data~\cite{Linde_Handbook}, educational data~\cite{LanWSB_Sparse}, the data generated by recommendation systems~\cite{GoldbNOT_Using,Srebr_Learning,RenniS_Fast}, and others;
\item {\bf machine learning}: low-rank models are ubiquitous in machine learning, laying the foundation for both classical techniques such as principal component analysis~\cite{Pears_Lines,Hotel_Analysis} as well as modern approaches to multi-task learning~\cite{ArgyrEP_Convex,OboziTJ_Joint} and natural language processing~\cite{Dumai_Latent,Blei_Probabilistic};
\item {\bf quantum state tomography}: a pure quantum state of $N$ ions can be described by a $2^N \times 2^N$ matrix with rank one~\cite{GrossLFBE_Quantum}.
\end{itemize}

In all of these settings, the matrices we are ultimately interested in can be extremely large.  Moreover, as data becomes increasingly cheap to acquire, the potential size will continue to grow.  This raises a number of challenges, but often a key obstacle is that fully observing the matrix of interest can prove to be an impossible task: it can be prohibitively expensive to fully sample the entire output of a sensor array; we might only be able to measure the strength of a few connections in a graph; and any particular user of a recommendation system will provide only a few ratings. In such settings we are left with a highly incomplete set of observations, and unfortunately, many of the most popular approaches to processing the data in the applications where low-rank matrices arise assume that we have a fully-sampled data set and are generally not robust to missing/incomplete data.  In these situations we are confronted with the inverse problem of recovering the full matrix from our incomplete observations.

While such recovery is not always possible in general, when the matrix is low rank, it is possible to exploit this structure and to perform this kind of recovery in a surprisingly efficient manner.  In fact, in recent years there has been tremendous progress in our understanding of how to solve such problems.  While many of these applications have a relatively long history in which various existing approaches to dealing with incomplete observations have been independently developed, recent advances in the closely related field of compressive sensing~\cite{Donoh_Compressed,Cande_Compressive,DavenDEK_Introduction} have enabled a burst of progress in the last few years.  We now have a unified framework which provides a strong base of theoretical results concerning when it is possible to recover a low-rank matrix from incomplete observations using efficient, practical algorithms~\cite{Gross_Recovering,CandeR_Exact,CandeT_Power,KeshaMO_Matrix,Recht_Simpler,RechtFP_Guaranteed}.

In this paper we provide a survey of this developing field.  We begin with a more formal mathematical statement of the problem of low-rank matrix recovery in Section~\ref{sec:mrprob}, followed in Section~\ref{sec:mralgs} by an overview of some of the algorithms most commonly used in practice to solve these kinds of problems.  We provide a brief overview of the existing theoretical guarantees for these algorithms in Sections~\ref{sec:gauss},~\ref{sec:mc}, and~\ref{sec:nonlinear} for several concrete observation models with an emphasis on how many observations are required to reliably recover a low-rank matrix and what additional assumptions are potentially required. Finally, in Section~\ref{sec:lifting} we describe an important application of these techniques to an important class of problems where we can solve {\em quadratic} and {\em bilinear} systems of equations by re-casting them as a simple problem of low-rank matrix recovery.

%----------------------------------------------------------------------------------
\section{The Matrix Recovery Problem}
\label{sec:mrprob}

We begin by carefully stating what we mean by {\em low-rank matrix recovery}.
%Low-rank matrices arise in a tremendous variety of scientific and engineering settings.  Practical applications where low-rank matrices arise include principle components analysis and multidimensional scaling~\cite{Pears_Lines,Hotel_Analysis,BorgG_Modern,CoxC_Multidimensional}, recommendation systems~\cite{GoldbNOT_Using,Srebr_Learning,RenniS_Fast}, rank aggregation~\cite{GleicL_Rank}, low-order system identification and blind deconvolution~\cite{Parti_Introduction,LiuV_Interior,AhmedRR_Blind}, array processing and sensor networks~\cite{BaronDSWB_Distributed,DavieE_Rank}, sensor localization~\cite{BiswaLWY_Semidefinite,LiniaLR_Geometry,SoY_Theory}, nuclear magnetic resonance spectroscopy~\cite{Singe_Remark,SingeC_Uniqueness}, quantum state tomography~\cite{Gross_Recovering,GrossLFBE_Quantum}, and many others.
We observe a matrix $\target$, which we will assume to have size $\M \times \N$ and which we can express either exactly or approximately as having rank $\R \ll \M, \N$.  This means that we can write
\[
	\target \approx \sum_{k=1}^\R \sigma_k \bu_k \bv_k^\T
\]
for some scalars $\sigma_1, \sigma_2, \ldots, \sigma_\R \ge 0$ and orthonormal vectors $\bu_1, \bu_2, \ldots, \bu_\R \in \Real^{\M}$ and $\bv_1, \bv_2, \ldots, \bv_\R \in \Real^{\N}$.  The $\{\sigma_k\}$ can be interpreted as the $R$ largest singular values of $\target$, and the $\{\vu_k\},\{\vv_k\}$ as the corresponding singular vectors.  The collection of all such matrices form a {\em union of subspaces} in $\Real^{\M\times\N}$; each set of vectors $\{\vu_k\},\{\vv_k\}$ define an $\R$-dimensional subspace, and the $\{\sigma_k\}$ correspond to an expansion in that subspace.  As the $\{\vu_k\},\{\vv_k\}$ can vary continuously, this union contains uncountably many such subspaces.
%where $\sigma_1, \sigma_2, \ldots, \sigma_\R \ge 0$ are the $\R$ largest singular values, and $\bu_1, \bu_2, \ldots, \bu_\R \in \Real^{\M}$, $\bv_1, \bv_2, \ldots, \bv_\R \in \Real^{\N}$ are the singular vectors.
%One can easily observe
However, by counting the degrees of freedom in the singular value decomposition, we see that the set of rank $\R$ matrices can be parameterized by $O(\R(\M+\N))$ numbers, which when $\R$ is small is much less than the $\M \N$ required to specify a general matrix.  This suggests that it might be possible to accurately recover a low-rank matrix from relatively few measurements, and as is noted above, recent results have shown that this is indeed possible~\cite{Gross_Recovering,CandeR_Exact,CandeT_Power,KeshaMO_Matrix,Recht_Simpler,RechtFP_Guaranteed}.

We will assume that rather than observing $\target$ directly we instead observe $\by = \cA(\target) + \bz$ where $\bz$ represents noise and $\cA: \Real^{\M \times \N} \rightarrow \Real^\obs$ is a linear {\em measurement operator} that acts on a matrix $\target$ by taking standard inner products against $\obs$ pre-defined $\M\times\N$ matrices $\mA_1,\ldots,\mA_\obs$:
\begin{equation} \label{def:A}
\twoCol{
\begin{aligned}
y_i  = \langle \target, \bA_i  \rangle + z_i  & = \trace(\bA_i^\T\target) + z_i \\
& = \sum_{m=1}^\M\sum_{n=1}^\N X_0[m,n]A_i[m,n] + z_i.
\end{aligned}
}
{	y_i = \langle \target, \bA_i  \rangle + z_i = \trace(\bA_i^\T\target) + z_i = \sum_{m=1}^\M\sum_{n=1}^\N X_0[m,n]A_i[m,n] + z_i.}
\end{equation}
%Here, the $\bA_i$ are $\M \times \N$ matrices and $\langle \cdot , \cdot \rangle$ denotes the usual matrix inner product.
Our discussion below will at times involve the adjoint of this operator, which is defined as
\[
	\cA^*(\vw) = \sum_{i=1}^\obs w_i\mA_i.
\]

Our survey focuses on three basic variations of this measurement model.  The first is taking $\cA$ to be a {\em random projection}, where each of the $\mA_i$ consisting of independent and identically distributed random variables.  Although this model arises in only a limited number of practical situations, the theory is so streamlined that it can be understood almost from first principles (see Section~\ref{sec:gauss}).  For our second model, $\cA$ returns a subset of the entries of the target.  Recovering from these samples is known as the {\em matrix completion} problem.  In this case, each of the $\mA_i$ has exactly one non-zero entry.  The analysis for this problem, which we overview in Section~\ref{sec:mc}, can also be extended to observing a subset of the expansion coefficients of $\target$ in a fixed (and known) orthobasis.  The third model, which we will discuss in Section~\ref{sec:lifting}, is that the $\mA_i$ are rank-1 matrices.  These are encountered when each observation can be written as a quadratic or bilinear form in $\target$.

While these are the measurement models that have received the most attention in the literature, they are by no means the only interesting models.  Other models inspired by applications in imaging and signal processing have also appeared recently in the literature (see for example \cite{krahmer14st,ahmed15co,ahmed15co2,tang14co,bahmani15li}).

%---
%In a typical matrix completion scenario, the $\bA_i$ will consist of matrices with only a single nonzero value of $1$ corresponding to the row and column of the observed element of $\target$, but in some cases the measurements may depend on multiple elements of $\target$, in which case the $\bA_i$ may have multiple nonzeros.  In either case, the goal will be to recover the low-rank matrix $\target$ from the measurements $\by$.

%----
%To recover $\target$, we might try to find a low-rank matrix $\bX$ such that $\cA(\bX)$ is as close as possible to the measurements $\by$, i.e.,
%\begin{equation}\label{eqn:rankmin}
%\widehat{\bX} = \argmin_{\bX \in \R^{\M \times \N}} \| \cA(\bX) - \by \|_2 \qquad \text{subject to} \qquad \rank(\bX) \le \R.
%\end{equation}
%Unfortunately, the low-rank constraint is not convex and so the resulting optimization problem is computationally intractable.

%----------------------------------------------------------------------------------
\section{Algorithms for Matrix Recovery}
\label{sec:mralgs}

\subsection{Low-rank approximation}

We start by reviewing the classical problem of finding the best low-rank approximation to a given $\M\times\N$ matrix $\mX_0$.  By ``best'', we mean closest in the sum-of-squares sense, and we formulate the problem as
\begin{equation}
	\label{eq:lrapprox}
	\minimize_{\mX}\|\mX-\mX_0\|_F^2\quad\text{subject to}\quad \rank(\mX) = \R,
\end{equation}
where $\|\mX-\mX_0\|_F^2 = \sum_{m,n}(X[m,n]-X_0[m,n])^2$ is the square of the standard Frobenius norm, and $\R$ is the desired rank of the approximation.  This problem is nonconvex, but is actually easy to solve explicitly using the {\em singular value decomposition} (SVD).  In particular, if we decompose $\mX_0$ as
\[
	\mX_0 = \mU\mSigma\mV^\T = \sum_{k=1}^{K}\sigma_k\vu_k\vv_k^\T,
\]
where $K=\min(\M,\N)$, $\mU,\mV$ are $\M\times K$ and $\N\times K$ matrices with orthonormal columns $\vu_1,\ldots,\vu_K$ and $\vv_1,\ldots,\vv_K$, and $\mSigma$ is a diagonal $K\times K$ matrix with sorted entries $\sigma_1\geq\sigma_2\geq\cdots\geq\sigma_K\geq 0$, then the solution to \eqref{eq:lrapprox} is found simply by truncating this expansion:
\[
	\widehat{\mX} = \sum_{k=1}^{R}\sigma_k\vu_k\vv_k^\T.
\]
This is known as the Eckart-Young theorem; see \cite[Chapter 7]{horn85ma} for a detailed proof and discussion.
For medium scale problems, computing the SVD to high precision is tractable, with computational complexity scaling as $O(K^2\max(\M,N))$.

Computing the best low-rank approximation, then, is akin to {\em thresholding} the singular values: we take the matrix, compute its SVD, keep the large singular values while killing off the small ones, and then reconstruct.  A variation of the program above makes this connection clearer.  The Lagrangian of \eqref{eq:lrapprox} is
\begin{equation}
	\label{eq:lrapproxlagrange}
	\minimize_{\mX}\|\mX-\mX_0\|_F^2 + \lambda\cdot\rank(\mX).
\end{equation}
As we vary the parameter $\lambda$ above, the solution to the program changes --- in fact, the set of solutions produced for different $0\leq\lambda<\infty$ is exactly the same as the set of solutions for \eqref{eq:lrapprox} produced for all $1\leq\R\leq K$.  Given $\lambda$, we solve \eqref{eq:lrapproxlagrange} by computing the SVD, hard thresholding the singular values via
\begin{equation}
	\label{eq:hardthreshold}
	\sigma_k' = \begin{cases} \sigma_k, & \sigma_k\geq \gamma, \\ 0, & \sigma_k < \gamma, \end{cases}
\end{equation}
with $\gamma = \sqrt{\lambda}$, and then taking $\widehat{\mX} = \mU\mSigma'\mV^\T$.

A common variation of the algorithm above involves replacing the hard threshold in \eqref{eq:hardthreshold} with a {\em soft} threshold.  In this case we still set the singular values that are small to zero, but now the large values are shrunk:
\begin{equation}
	\label{eq:softthreshold}
	\sigma_k' = \begin{cases} \sigma_k-\gamma, & \sigma_k\geq\gamma, \\ 0, & \sigma_k < \gamma. \end{cases}
\end{equation}
This amounts to a more gradual phasing out of the terms that just cross the threshold.  It turns out that this soft thresholding process can also be put in variational form; when $\gamma = \lambda/2$ the result of the procedure above is the solution to
\begin{equation}
	\label{eq:nnlagrange}
	\minimize_{\mX} \|\mX-\mX_0\|_F^2 + \lambda\|\mX\|_*,
\end{equation}
where $\|\mX\|_*$ is the {\em nuclear norm}, and is equal to the sum of the singular values of $\mX$.
%
%\edit{Mention ``trace norm'' here? Maybe worth noting that the reason for considering this version is that~\eqref{eq:nnlagrange} is convex?}
%
$\|\mX\|_*$ is also known as the {\em trace norm}, as it is equal to the trace when $\mX$ is symmetric positive semidefinite.  Unlike the rank, $\|\mX\|_*$ is a convex function, and often appears as a convex proxy for rank in optimization problems \cite{fazel02phd}.  While in the approximation problem we are considering here, the solutions to \eqref{eq:lrapproxlagrange} and \eqref{eq:nnlagrange} involve very similar computations, this will not be the case at all when we consider recovery from partial observations in the next section.

In the procedures above, the computational cost is dominated by computing the SVD.  For matrices with $\M,\N$ on the order of $100$--$1000$, there are a number of exact methods with similar computational complexity --- see \cite[Chap.\ 45]{hogben13ha} for an overview.  When $\mX_0$ is large but is very well approximated by a matrix with modest rank, randomized algorithms can be used to compute an approximate SVD \cite{talwalkar13la,halko11fi,mahoney11ra}.

%-----------------------------------------------------------------------------
\subsection{Low-rank recovery and nuclear norm minimization}

The low-rank {\em approximation} problem described above readily admits a straightforward solution.  However, in this survey we are concerned instead with the low-rank {\em recovery} problem where we are working from (possibly noisy) indirect observations, $\vy\approx\cA(\mX_0)$.  In this case we would ideally like to solve the analog of \eqref{eq:lrapprox},
\begin{equation*} \label{eq:lrrecovery}
	\minimize_{\mX} \|\vy-\cA(\mX)\|_2^2 \quad\text{subject to}\quad \rank(\mX)=\R.
\end{equation*}
Unfortunately, whereas~\eqref{eq:lrapprox} could be solved via a simple SVD,~\eqref{eq:lrrecovery} is in general NP-hard.

In contrast, the nuclear norm minimization program remains tractable. In particular, with indirect observations, we replace \eqref{eq:nnlagrange} with
\begin{equation}
	\label{eq:matrixlasso}
	\minimize_{\mX} \|\cA(\mX)-\vy\|_2^2 + \lambda\|\mX\|_*.
\end{equation}
This is an unconstrained convex optimization program, and can be solved in a systematic way using a {\em proximal algorithm} \cite{parikh13pr,combettes11pr}.  The solution(s) to \eqref{eq:matrixlasso} will obey, for any $\gamma>0$, the fixed point condition \cite{combettes11pr}
\[
	\mX_\star = \operatorname{prox}_{\gamma}(\mX_\star - \gamma\cA^*(\cA(\mX_\star) - \vy )
\]	
where the proximal operator is
\begin{equation}
	\label{eq:nnprox}
	\operatorname{prox}_{\gamma}(\mZ) = \argmin_{\mX} \|\mX-\mZ\|_F^2 + \gamma\lambda\|\mX\|_*.
\end{equation}
One class of methods for solving \eqref{eq:matrixlasso} are based on the iteration
\begin{equation}
	\label{eq:proxiter}
	\mX_{k+1} = \operatorname{prox}_{\gamma_k}(\mX_k - \gamma_k\cA^*(\cA(\mX_k) - \vy ),
\end{equation}
for some appropriately chosen sequence $\{\gamma_k\}$ \cite{cai10si}.  As discussed in the previous section, the subproblem \eqref{eq:nnprox} is solved by singular value soft-thresholding.  Computing the SVD of the $\mX_k - \gamma_k\cA^*(\cA(\mX_k) - \vy )$ at each iteration is almost always the dominant cost, as it typically requires significantly more computation than applying $\cA$ and $\cA^*$.

State-of-the-art methods for solving \eqref{eq:matrixlasso} based on singular value thresholding are not too much more complicated.  For example, the FISTA algorithm \cite{beck09fa} modifies the basic iteration in \eqref{eq:proxiter} through intelligent choices of the scaling coefficient $\gamma_k$ and by replacing $\mX_k$ with a carefully chosen combination of $\mX_k$ and $\mX_{k-1}$.  These small changes have almost no effect on the amount of computation done at each iteration, but they converge in significantly fewer iterations.

%\vspace{.25in}

The simplicity of these proximal-type algorithms makes them very attractive for small to medium sized problems.  However, as the number of rows and columns in the matrix gets to be several thousand, direct computation of the SVD becomes problematic.  For specially structured $\cA$, including the important case where $\cA(\mX)$ returns a subset of the entries of $\mX$, fast algorithms that take advantage of this structure to compute the SVD have been developed to solve \eqref{eq:matrixlasso} or closely related problems \cite{mazumder10sp,cai10si}.  In more general settings, techniques from randomized linear algebra have been applied to compute approximate SVDs \cite{zhou11go}.

Storage is also an issue when the target matrix is large.  Since we expect the target matrix to have small rank, we would like to save on storage by restricting the iterates to also be low rank.  We can reformulate the program above with  $\mX\approx \mL\mR^\T$, and optimize over the $\M\times\R$ and $\N\times\R$ matrices $\mL$ and $\mR$ rather than the $\M\times\N$ matrix $\mX$.  This reformulation is driven by the fact that the nuclear norm is equal to the minimum Frobenius norm factorization \cite{RechtFP_Guaranteed}:
\[
	\|\mX\|_* = \min_{\mL,\mR} \frac{1}{2}\left(\|\mL\|_F^2 + \|\mR\|_F^2\right)
	\quad\text{subject to}\quad \mX=\mL\mR^\T.
\]
We can then replace \eqref{eq:matrixlasso} with
\begin{equation}
	\label{eq:bmnonconvex}
	\minimize_{\mL,\mR} \|\cA(\mL\mR^\T)-\vy\|_2^2 + \frac{\lambda}{2}\|\mL\|_F^2 + \frac{\lambda}{2}\|\mR\|_F^2,
\end{equation}
and if the solution to \eqref{eq:matrixlasso} does indeed have rank at most $\R$, it will also be the solution to \eqref{eq:bmnonconvex}.  While this new formulation is non-convex --- $\cA(\mL\mR^\T)$ is a combination of products of unknowns --- there are assurances that the local minima in \eqref{eq:bmnonconvex} are also global minima if the rank of the true solution is smaller than $\R$ \cite{burer05lo}.
% \cite{bach} above, too, but I couldn't find the reference ...
%
This technique is sometimes referred to as the {\em Burer-Monteiro heuristic}, after the authors of \cite{burer03no,burer05lo} who proposed a version of the above for general semidefinite programming, and it is used in state-of-the-art large scale implementations of matrix recovery problems \cite{recht13pa}.

The parameter $\lambda$ in \eqref{eq:matrixlasso} determines the trade-off between the fidelity of the solution to the measurements $\vy$ and its conformance to the low-rank model.  When we are very confident in the measurements, it might make sense to use them to define a set of linear equality constraints, solving
\begin{equation}
	\label{eq:nneq}
	\minimize_{\mX}~\|\mX\|_*\quad\text{subeject to}\quad \cA(\mX)=\vy.
\end{equation}	
The output of this program will match \eqref{eq:matrixlasso} as $\lambda\rightarrow 0$.  Some of the analytical results we review in Sections~\ref{sec:gauss} and \ref{sec:mc} reveal conditions under which \eqref{eq:nneq} recovers a low-rank matrix $\target$ exactly given measurements $\vy=\cA(\mX_0)$ as constraints.

%-----------------------------------------------------------------------------
\subsection{Iterative hard thresholding}

From the algorithmic point of view, iterative hard thresholding (IHT) algorithms \cite{jain10gu,tanner13no} are very similar to the proximal algorithms used to solve nuclear norm minimization in the previous section.  However, when the target is very low rank, they tend to converge extremely quickly.

The basic iteration is as follows.  From the current estimate $\mX_k$, we first take a step in the direction of the gradient of $\|\cA(\mX)-\vy\|_2^2$, then project onto the set of rank $\R$ matrices:
\begin{align*}
	\mY_{k+1} &= \mX_k - \gamma_k\cA^*(\cA(\mX_k) - \vy)) \\
	\mX_{k+1} &= \operatorname{ProjectRank}_{\R}(\mY_{k+1}).
\end{align*}
The $\operatorname{ProjectRank}_{\R}$ operator computes the top $\R$ left and right singular vectors and singular values --- when $\R$ is small compared to $\M$ and $\N$, this can be done in significantly less time than computing a full SVD \cite{simon00lo}, especially if the operator $\cA$ and its adjoint $\cA^*$ are structured in a such a way that there is a fast method for applying the matrices $\mY_{k+1}$ to a series of vectors.  In these cases, the intermediate matrix $\mY_{k+1}$ is not computed explicitly, but each term in the first equation above can be handled efficiently in the SVD computation.  The implementations of IHT in \cite{jain10gu,tanner13no} rely on existing software packages \cite{propack} to do this.

In contrast to the nuclear norm minimization algorithm above, each of the iterates $\mX_k$ IHT produces has a prescribed rank.  The storage required for $\mX_k$ is roughly $\R(\M+\N)$, as opposed to the $\M\N$ required for a general $\M\times\N$ matrix.  This difference is critical for large-scale applications.

% PROPACK paper
% http://sun.stanford.edu/~rmunk/PROPACK/paper.pdf

%-----------------------------------------------------------------------------
\subsection{Alternating projections}
\label{sec:alternatingprojections}

The alternating projections algorithm  is another space efficient technique which stores the iterates in factored form.  The algorithm is extraordinarily simple, and easy to interpret: looking for a $\M\times\N$ rank $\R$ matrix that is consistent with $\vy$:
\[
	\minimize_{\mX}~\|\cA(\mX)-\vy\|_2^2,\quad\text{subject to}\quad\rank(\mX)=\R,
\]
is the same as looking for a $\M\times\R$ matrix $\mL$ and a $\N\times\R$ matrix $\mR$ whose product is consistent with $\vy$:
\begin{equation}
	\label{eq:altminopt}
	\minimize_{\mL,\mR}~\|\cA(\mL\mR^\T)-\vy\|_2^2.
\end{equation}
This optimization problem is still non-convex, but with one of $\mL$ or $\mR$ fixed, it is a simple least-squares problem.  This motivates the following iteration.  Given current estimates $\mL_k$,$\mR_k$, we update using
\begin{equation}
	\label{eq:altminiter}
    \begin{aligned}
	\mR_{k+1} &= \arg\min_{\mR}\|\cA(\mL_k\mR^\T) - \vy\|_2^2, \\
	\mL_{k+1} &= \arg\min_{\mL}\|\cA(\mL\mR_{k+1}^\T) - \vy\|_2^2.
    \end{aligned}
\end{equation}
Each step involves solving a linear system of equations with $\R\M$ or $\R\N$ variables for which we can draw on well-established algorithms in numerical linear algebra.  Its simplicity and efficiency make it one of the most popular methods for large-scale matrix factorization \cite{koren09ma}, and it tends to outperform nuclear norm minimization \cite{haldar09ra,jain13lo}, especially in cases where the rank $\R$ is very small compared to $\N,\M$.

There are few general convergence guarantees for alternating projections, and the final solution tends to depend heavily on the initialization of $\mL$ and $\mR$.  However, guarantees for the rate of convergence can be found in \cite{Kesha_Efficient} and recent work \cite{jain13lo} has provided some first theoretical results for conditions under which the iterations above converge to the true low-rank matrix (and a method for supplying a reliable starting point).  These are discussed further in Sections~\ref{sec:gauss} and \ref{sec:mc} below.

Another advantage alternating projections is that the framework can be extended to handle structure on one or both of the factors $\mL$ and $\mR$.  Typically, this means that the least-squares problems above are either regularized or constrained in a manner which encourages or enforces the desired structure.  Extensions of the iterations above have been used successfully for problems including non-negative matrix factorization \cite{kim08no}, sparse PCA, where we restrict the number of nonzero terms in  $\mL,\mR$ \cite{zou06sp}, and dictionary learning \cite{olshausen96na,lewicki99pr}, where $\mL$ is a well-conditioned matrix and $\mR$ is sparse.  Again, there are few strong theoretical guarantees for these algorithms, with notable exceptions in the recent works \cite{lee13ne,agarwal14le}.

The optimization problems for alternating projections \eqref{eq:altminopt} and the Burer-Monteiro heurtistic \eqref{eq:bmnonconvex} are similar (nonlinear) least-squares problems on the matrix factors $\mL,\mR$.  The algorithms used to solve them, though, have a distinct difference: instead of fixing for one factor and optimizing the other as in \eqref{eq:altminiter} above, solvers for \eqref{eq:bmnonconvex} (e.g.,~\cite{recht13pa}) typically take descent steps on $\mL$ and $\mR$ simultaneously.  Convergence analysis for closely related local descent methods can be found in the recent works \cite{TuBSR_Lowrank,BhojKS_Dropping}.

% Moritz Hardt
% http://arxiv.org/abs/1312.0925

%-----------------------------------------------------------------------------
\subsection{Other algorithms for matrix recovery}

The methods above are by no means the only algorithms which have been proposed for low-rank matrix recovery.  We close this section by briefly mentioning some other techniques.

Recent years have seen a renewed interest in Frank-Wolfe-type algorithms for minimizing norms defined by the convex hull of a set of atoms~\cite{clarkson10co,jaggi13re}.  These algorithms are of particular interest for minimizing the nuclear norm (where the atoms are rank-1 matrices), as they only require computing the leading singular vector at every iteration, rather than a full SVD~\cite{jaggi10si}.
The nuclear norm problems in~\eqref{eq:matrixlasso} and~\eqref{eq:nneq} can also be minimized by solving a series of weighted least-squares problems, each of which can be solved using standard techniques for linear systems of equations~\cite{fornasier11lo,MohanF_Iterative,FornaPRW_Conjugate}.

Beyond the nuclear norm, other proxies for rank exist.  The max-norm~\cite{srebro04ma,srebro05ra} is a convex function that results in a similar optimization program to~\eqref{eq:matrixlasso}, and is subject to similar heuristics as~\eqref{eq:bmnonconvex} for storage reduction~\cite{recht13pa}.  Alternatively, the logarithm of the determinant is a nonconvex, but smooth, proxy for rank when $\mX$ is positive semidefinite (and can be applied to general matrices by embedding them in a PSD matrix).  In practice, locally minimizing this function subject to convex constraints tends to produce low-rank solutions~\cite{FazelHB_Logdet}.

A greedy algorithm for low-rank recovery was presented in~\cite{lee10ad}.  This algorithm alternates between selecting an estimate for the $\R$-dimensional subspace in which $\target$ lives, and projecting onto these subspaces; it is equipped with strong theoretical guarantees.  Another class of algorithms evolves the left and right singular vectors along the Grassman manifold~\cite{dai11su,KeshaMO_Matrix}.  These algorithms, which are specialized for the matrix completion problem, are computationally efficient, and have equally strong theoretical performance guarantees.

%----------------------------------------------------------------------------------
\section{Matrix Recovery from Gaussian Observations}
\label{sec:gauss}

The theory of low-rank recovery is clean and elegant when the measurement operator $\cA(\cdot)$ is a random projection.  Applications where this is a good model for the observations are limited, but looking at it as an abstract problem gives us real mathematical insight about why low-rank recovery works.

The discussion in this section will center on ``Gaussian'' $\cA(\cdot)$, where the entries of each of the $\mA_m$ are independent and identically distributed normal random variables with zero mean and variance $L^{-1}$ --- this variance is chosen so that $\E[\|\cA(\mX)\|_2^2] = \|\mX\|_F^2$ for any fixed $\M\times\N$ matrix $\mX$.

\subsection{The matrix restricted isometry property}
\label{ssec:MRIP}

We first examine the fundamental question of whether we can distinguish different rank-$\R$ matrices viewed through the lens of the operator $\cA$.  One way to formalize this is asking whether $\cA$ preserves the distances between all such matrices; this is certainly true if there exists a $0\leq\delta<1$ such that
\begin{equation}
	\label{eq:matrix-rip}
\twoCol{1-\delta
	~\leq~
	\frac{\|\cA(\mX_1) - \cA(\mX_2)\|_2^2}{\|\mX_1-\mX_2\|_F^2}
	~\leq~
	1+\delta}
{	(1-\delta)\|\mX_1-\mX_2\|_F^2
	~\leq~
	\|\cA(\mX_1) - \cA(\mX_2)\|_2^2
	~\leq~
	(1+\delta)\|\mX_1-\mX_2\|_F^2}
\end{equation}
for all $\mX_1,\mX_2$ of rank $\R$ or smaller.  This condition is known as the {\em matrix restricted isometry property} (matrix-RIP), and is the matrix analog to the restricted isometry property from compressive sensing~\cite{DavenDEK_Introduction}. The first immediate consequence is that all matrices of rank $\R$ or less have unique images, since if $\cA(\mX_1)=\cA(\mX_2)$ for $\mX_1\not=\mX_2$, then the lower bound would be violated.  Qualitatively, the upper and lower bounds tell us that two rank $\R$ matrices are as distinguishable from their measurements as they would be if they were observed directly.  With \eqref{eq:matrix-rip} established, we can use any number of techniques to recover a low-rank matrix from measurements through $\cA$; we discuss some specific guarantees below.

A Gaussian $\cA$ will almost certainly have the  matrix restricted isometry property when the number of observations are commensurate with the number of degrees of freedom for an $\M\times\N$ matrix with rank $\R$.  Specifically, \eqref{eq:matrix-rip} holds with high probability when\footnote{We use this compact notation to mean that there is a constant so that $\obs\geq\mathrm{Const}\cdot\R(\M+\N)$.}
\begin{equation}
	\label{eq:gaussianmeas}
	\obs ~\gtrsim~ \R(\N+\M).
\end{equation}
This can be established through relatively simple probabilistic methods.  We sketch the argument below; the detailed proof in~\cite{CandeP_Tight} smartly combines ideas from~\cite{RechtFP_Guaranteed} and~\cite{Versh_Introduction}.

 To begin, notice that it is enough to show that $(1-\delta)\leq\|\cA(\mX)\|_2^2\leq(1+\delta)$ for all $\mX$ of rank $2\R$ and unit Frobenius norm.  Then there are three basic steps for establishing \eqref{eq:gaussianmeas}.
\begin{enumerate}

	\item For any arbitrary fixed $\M\times\N$ matrix $\mX$ with $\|\mX\|_F^2=1$,
	\begin{equation}
		\label{eq:tailbound}
		\P{\left|\|\cA(\mX)\|_2^2 - 1\right| > t}
		~\leq~
		C\e^{-c\obs t^2}%,\quad\text{for $t<1/2$},
	\end{equation}
	for $t \le \frac12$, where $C$ and $c$ are reasonable constants that can be calculated explicitly (standard calculations yield $C\leq 2$ and $c\geq 1/8$).  Since the entries of $\cA(\mX)$ are independent Gaussian random variables, $\|\cA(\mX)\|_2^2$ is a chi-squared random variable with $\obs$ degrees of freedom, and the inequality above follows from standard tail bounds \cite{laurent00ad}.
	
	We want \eqref{eq:tailbound} to hold not just for a single matrix, but uniformly over the set of all unit-norm rank $2\R$ matrices.  It is straightforward to get a uniform result over any finite set $\setQ$ of such matrices by using a union bound:
	\[
		\P{\sup_{\mX\in\setQ}\left|\|\cA(\mX)\|_2^2 - 1\right| > t}
		~\leq~
		|\setQ|\cdot C\,\e^{-c\obs t^2}.
	\]
	Even though the set of all unit-norm rank $2\R$ matrices is infinite, the next step shows us that it is enough to consider a finite subset.
	
	\item Let $\setR$ denote the set of $\M\times\N$ matrices $\mX$ with $\rank(\mX)\leq\R$ and $\|\mX\|_F=1$.  Now let $\setR_\epsilon$ be any finite $\epsilon$-approximation to $\setR$ --- this means that $\setR_\epsilon\subset\setR$ and for every $\mX\in\setR$ there is an $\bar{\mX}$ in $\setR_\epsilon$ that is within $\epsilon$: $\|\mX-\bar{\mX}\|_F\leq\epsilon$.  Then a short argument shows that for $\epsilon = \delta/(4\sqrt{2})$,
    \[
		\max_{\mX\in\setR_\epsilon}\left|\|\cA(\mX)\|_2^2 - 1\right| \leq \delta/2
	\]
    implies that 
    \[
        \sup_{\mX\in\setR}\left|\|\cA(\mX)\|_2^2 - 1\right| \leq \delta.
	\]
	Thus the supremum over the infinite set on the right can be replaced with the maximum over the finite set on the left.  Using the result from step 1, we now have
	\[
		\P{\sup_{\mX\in\setR}\left|\|\cA(\mX)\|_2^2 - 1\right| > \delta}
		~\leq~|\setR_\epsilon|\cdot C\,\e^{-cL\delta^2/4}.
	\]
	
	\item The size of $\setR_\epsilon$ can be estimated as a function of $\epsilon$.  The bound in \cite{CandeP_Tight} reads
	\[
		|\setR_\epsilon| \leq \left(\frac{9}{\epsilon}\right)^{\R(\M+\N+1)} = ~\e^{\R(\M+\N+1)\log(9/\epsilon)}.
	\]
	Combining this with the choice of $\epsilon$ above and plugging it into the result of step 2, tells us that \eqref{eq:matrix-rip} will hold with probability at least $1-C'\e^{-c'\obs}$ when $\obs\gtrsim \delta^{-2}\R(\M+\N)$.
\end{enumerate}

Beyond basic identifiability, the matrix-RIP is also sufficient for concrete algorithms to produce accurate estimates of low-rank matrices.  For example, \cite{oymak11si,cai13co} refine an argument in \cite{RechtFP_Guaranteed} to show that nuclear norm minimization will recover rank $\R$ $\mX_0$ when $\delta\leq 0.3$ in \eqref{eq:matrix-rip}.  A similar result holds for iterative hard thresholding techniques \cite{jain10gu} and their variants \cite{tanner13no}: when $\delta\leq 1/3$, we are guaranteed to recover rank-$\R$ $\mX_0$ with linear convergence.  The best known guarantees for alternating minimization are weaker: \cite{jain13lo} states that $\delta\leq \mathrm{Const}/\R$ is sufficient for linear convergence, which is achieved for Gaussian $\cA$ when $\obs\gtrsim\R^3(\M+\N)$.

The uniformity of the matrix-RIP, that it holds for all pairs of rank $\R$ matrices simultaneously, results in stability guarantees for each of these algorithms when the measurements are made in the presence of noise, or the target matrix is only approximately low rank.

%-----------------------------------------------------------------------------
\subsection{Convex geometry and Gaussian widths}

One strength of the analysis described above (based on the matrix-RIP) is that it is quite general and can help in analyzing a variety of different algorithms.  However, the resulting bounds are often quite loose.  In the specific case of nuclear norm minimization, however, there are very precise conditions under which it will succeed (or fail) with high probability with a Gaussian measurement ensemble. These results are of a slightly different nature than those based on the matrix-RIP in the previous section.  They are weaker, in that they only speak to the recovery of a single matrix; the target $\mX_0$ is fixed, the ensemble $\{\mA_\ell~:~\ell=1\,\ldots,\obs\}$ is generated independently of $\mX_0$, and the probability with which $\mX_0$ is recovered is computed. But they are also stronger in the sense that they are much more precise in telling us when recovery will succeed or fail. In this section, we will focus on  results outlining the conditions for recovering $\mX_0$ perfectly in the absence of noise.  Given $\vy=\cA(\mX_0)$, we solve the equality constrained problem \eqref{eq:nneq}.
%\begin{equation}
%	\label{eq:nneq}
%	\minimize_{\mX} \|\mX\|_*\quad\text{subject to}\quad \cA(\mX) = \vy.
%\end{equation}

\begin{figure}
	\centering
\twoCol{
	\begin{tabular}{cc}
		\includegraphics[scale=0.3]{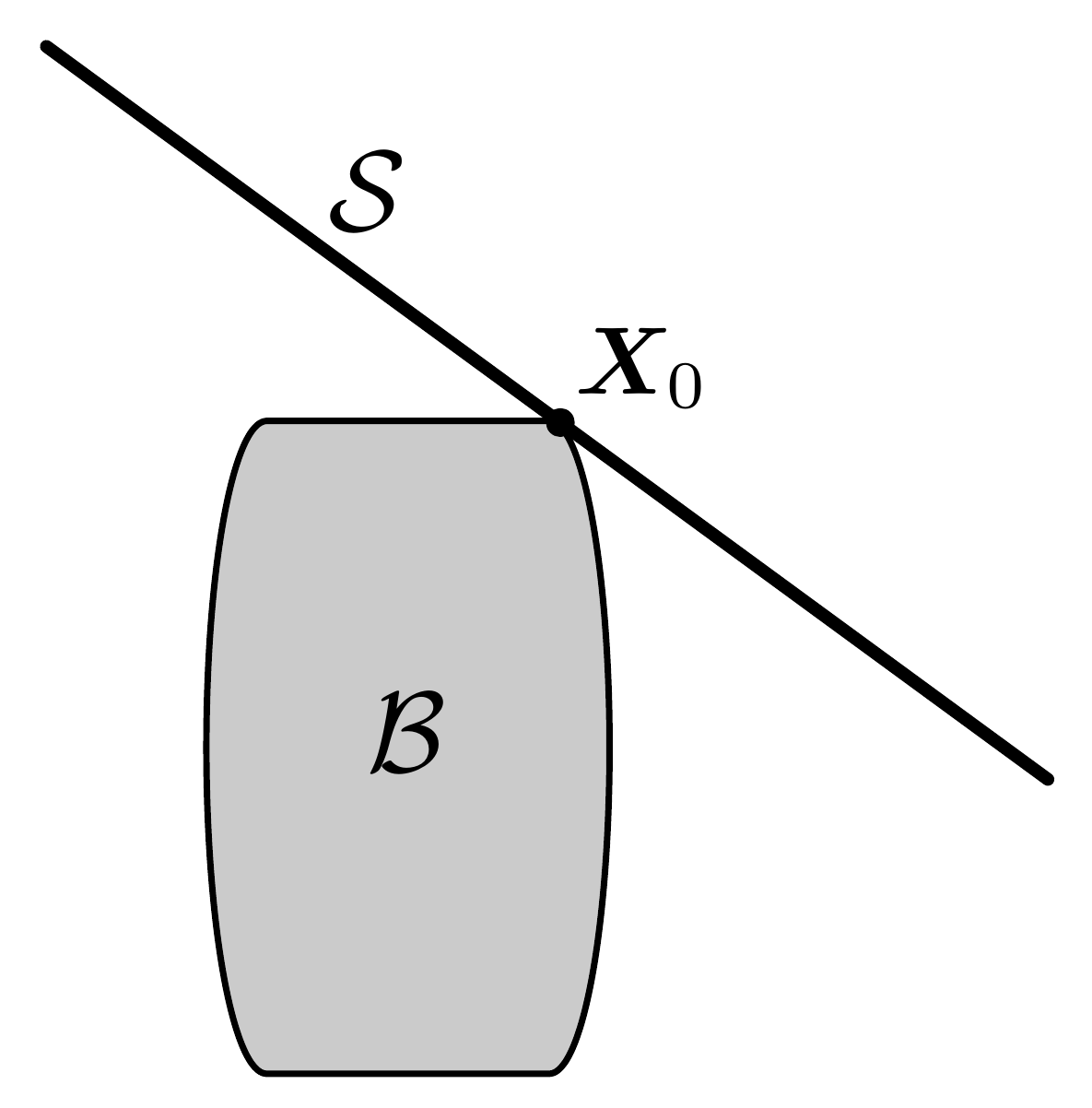} &
		\includegraphics[trim=50 110 100 80,clip,scale=0.6]{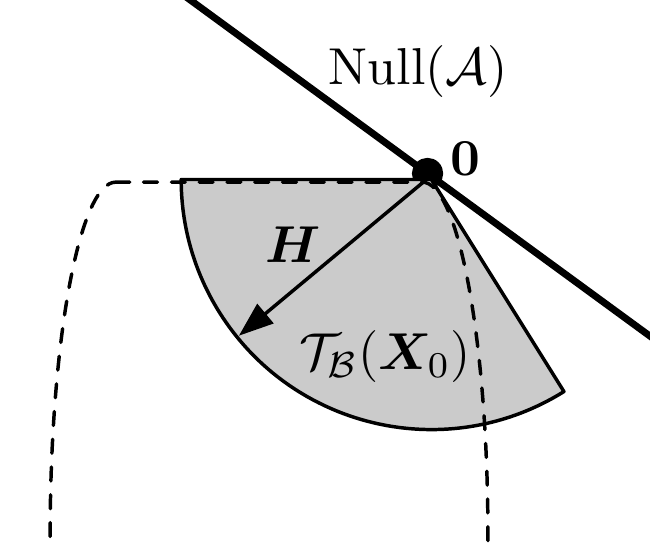} \\
		(a) & (b)
	\end{tabular}}
{	\begin{tabular}{ccc}
		\includegraphics[scale=0.3]{convgeom1} & \hspace{1in} &
		\includegraphics[trim=50 110 100 80,clip,scale=.6]{convgeom2} \\
		(a) & &  (b)
	\end{tabular}}
	\caption{\small\sl (a) The solution to the optimization program \eqref{eq:nneq} is $\mX_0$ when the norm ball $\setB$ defined in \eqref{eq:normball} and the affine space in \eqref{eq:affinespace} intersect at only one point.  (b) The tangent cone $\setT_{\setB}(\mX_0)$ from \eqref{eq:tangentcone} along with a representative element $\mH\in\setT_\setB(\mX_0)$. }
	\label{fig:ballplane}
\end{figure}

Recent works \cite{chandrasekaran12co,oymak11ti,amelunxen14li} analyze the performance of this program using very intuitive geometrical principles.  The target matrix $\mX_0$ is a member of two different convex sets; it is in the nuclear norm ball of radius $\|\mX_0\|_*$,
\begin{equation}
	\label{eq:normball}
	\mX_0 \in \setB := \{\mX~:~\|\mX\|_*\leq\|\mX_0\|_*\},
\end{equation}
and it is in the affine space consisting of all $\M\times\N$ matrices that have the same measurements,
\begin{equation}
	\label{eq:affinespace}
	\mX_0 \in \setS := \{\mX~:~\cA(\mX)=\vy\}.
\end{equation}
By definition, $\mX_0$ is the unique solution to \eqref{eq:nneq} if and only if it is the only matrix in the intersection of $\setB$ and $\setS$; this is illustrated in Figure~\ref{fig:ballplane}(a).  Generating $\cA$ with a Gaussian distribution is the same as choosing the orientation of $\setS$ uniformly at random.  The local geometry of the tip of $\setB$ is determined by the rank of $\mX_0$, smaller ranks make this point more singular, and hence decrease the probability of an intersection.  The dimension of the set $\setS$ is $\M\N-\obs$, the same as the null space of $\cA$; as $\obs$ increases, $\setS$ gets smaller, and the probability of an intersection decreases.

We can make both of these statements precise.  The collection of directions that lead into the ball (i.e., decrease the nuclear norm) $\setB$ from the point $\mX_0$ is called the {\em tangent cone} of $\setB$ at $\mX_0$:
\begin{equation}
	\label{eq:tangentcone}
	\setT_{\setB}(\mX_0) = \{\mH~:~\|\mX_0+\epsilon\mH\|_*\leq\|\mX_0\|_*~~\text{for some $\epsilon>0$}\}.
\end{equation}
Asking if there is a better feasible point in \eqref{eq:nneq} than $\mX_0$ is exactly the same as asking if the subspace $\setS-\mX_0 = \Null(\cA)$ (the affine set $\setS$ shifted to the origin) intersects the cone $\setT_{\setB}(\mX_0)$ at any place other than the origin.

%---
% trim is left,bottom,right,top
%---

When does a randomly chosen subspace intersect a cone only at the origin?  A clean answer is given in \cite{chandrasekaran12co}, which builds on the classic work \cite{gordon88mi}. This answer depends on the notion of the {\em Gaussian width} of a set $\setC$, which is defined as
\[
w(\setC) = \E\left[\sup_{\mX\in\setC,\|\mX\|_F=1} \inner{\mX,\mG}\right],
\]
where $\mG$ is a matrix whose entries are independent zero-mean Gaussian random variables with unit variance, and the expectation is taken with respect to $\mG$. The quantity $w(\setC)$ can be interpreted as the amount we expect a set $\setC$ to align with a randomly drawn vector. It is shown in~\cite{chandrasekaran12co} that a randomly chosen subspace will intersect a cone $\setC$ only at the origin with high probability if the codimension of the subspace is at least as large as the square of the Gaussian width:
\[
	\operatorname{codim}(\Null(\cA)) ~\geq~ w(\setT_{\setB}(\mX_0))^2 + 1.
\]
It is clear that for subsets of $\M\times\N$ matrices,  $\omega(\setC)\leq\sqrt{\M\N}$, and if $\setC$ is a $D$ dimensional subspace, a quick calculation shows that $w(\setC)=\sqrt{D}$.  The codimension of $\Null(\cA)$ is always $\obs$, so measuring the Gaussian width of the tangent cone gives an immediate sufficient condition on the number of measurements needed for accurate recovery.  The bound on the Gaussian width $w(\setT_{\setB}(\mX_0))^2 ~\leq~ 3\R(\M+\N-\R)$ gives us the sharp sufficient condition of
\begin{equation}
	\label{eq:cprw-tight}
	\obs \geq 3R(\M+\N-\R)+1.
\end{equation}

The relation in \eqref{eq:cprw-tight} is very close to being necessary as well.  In \cite{oymak11ti,amelunxen14li}, it is shown that if the number of observations is not too far below $w(\setT_{\setB}(\mX_0))^2$, then the probability that \eqref{eq:nneq} fails to recover $\mX_0$ is very close to $1$.  These papers contain an impressive suite of numerical experiments showing that the success or failure of equality-constrained nuclear norm minimization \eqref{eq:nneq} can be predicted accurately from the parameters $\M,\N,\R$.

It should be mentioned that the analysis in \cite{chandrasekaran12co} and \cite{amelunxen14li} applies to many different types of structured recovery problems based on convex optimization; the low-rank recovery results discussed above are an important special case.

% Similar tight bounds using duality are given by Candes and Recht, '13

%----------------------------------------------------------------------------------
\section{Matrix Completion}
\label{sec:mc}

As we have just seen, the theory of low-rank matrix recovery is particularly elegant in the case where the measurement operator $\cA(\cdot)$ is a (Gaussian) random projection operator.  While this provides some insight into the kind of behavior we can hope for in many applications, it is also far from representative of the type of observations we often encounter in practice.  In particular, in many settings of interest the $\mA_i$ are highly structured --- in the case of matrix {\em completion}, the $\mA_i$ will have only a single nonzero value of 1 corresponding to the row and column of the observed element.  An equivalent way to think about this type of measurement is that we only observe the entries of $\mX_0$ on a subset $\Omega$ of the complete set of entries.  This kind of measurement model arises in a variety of practical  settings.  For example, we might have a large number of questions we would like to potentially ask a number of users (as in a large-scale survey or recommendation system) but in practice we might only expect to receive responses to a few of these questions from any given user.  Similarly, in many large-scale graphs (such as graphs representing the strength of social connections or the distances between sensors or other items) we might only be able to measure/observe the strength of a few connections in a graph.

Even if the elements of $\Omega$ are chosen at random, this scenario has some significant differences from the case where the $\mA_i$ are Gaussian. It is clear that we cannot expect the theory developed using Gaussian widths to be of much use, since our observations are not Gaussian, but there is a more fundamental problem that arises in the case of matrix completion.

\subsection{Which matrices can be completed?}
\begin{figure}
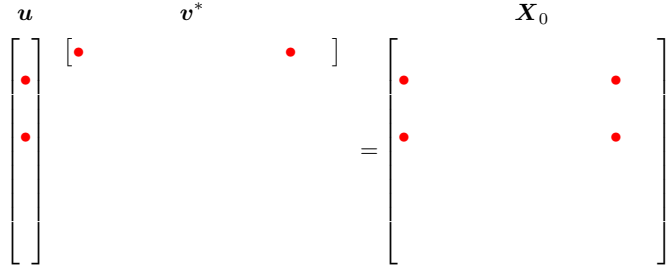

\centering
\resizebox{3.5in}{!}{\parbox{.5\linewidth}{%
\begin{align*}
\vu \hspace{0.86in} \vv^*\hspace{1.02in} &   \hspace{0.8in} \mX_0 \\
\begin{bmatrix}  \\ {\color{red} \bullet} \\  \\ {\color{red} \bullet} \\  \\   \\  \\ \\  \end{bmatrix}   \begin{array}{c} \begin{bmatrix}  {\color{red} \bullet}  &  \hphantom{0} & \hphantom{0}  & \hphantom{0}    &  \hphantom{0} & \hphantom{0}   & {\color{red} \bullet}  &  \hphantom{0}    \end{bmatrix}  \\ \\ \\ \\ \\  \\  \\ \\ \end{array}   = &
 \begin{bmatrix}
\hphantom{0} & \hphantom{0} & \hphantom{0} & \hphantom{0} & \hphantom{0} & \hphantom{0} & \hphantom{0} & \hphantom{0} \\
{\color{red} \bullet}  &  \hphantom{0} & \hphantom{0}  & \hphantom{0}    &  \hphantom{0} & \hphantom{0}   & {\color{red} \bullet}  &  \hphantom{0} \\
\hphantom{0} & \hphantom{0} & \hphantom{0} & \hphantom{0} & \hphantom{0} & \hphantom{0} & \hphantom{0} & \hphantom{0} \\
{\color{red} \bullet}  &  \hphantom{0} & \hphantom{0}  & \hphantom{0}    &  \hphantom{0} & \hphantom{0}   & {\color{red} \bullet}  &  \hphantom{0} \\
\hphantom{0} & \hphantom{0} & \hphantom{0} & \hphantom{0} & \hphantom{0} & \hphantom{0} & \hphantom{0} & \hphantom{0} \\
\hphantom{0} & \hphantom{0} & \hphantom{0} & \hphantom{0} & \hphantom{0} & \hphantom{0} & \hphantom{0} & \hphantom{0} \\
\hphantom{0} & \hphantom{0} & \hphantom{0} & \hphantom{0} & \hphantom{0} & \hphantom{0} & \hphantom{0} & \hphantom{0} \\
\hphantom{0} & \hphantom{0} & \hphantom{0} & \hphantom{0} & \hphantom{0} & \hphantom{0} & \hphantom{0} & \hphantom{0}
\end{bmatrix}
\end{align*}}}
    \caption{\small\sl  A low-rank matrix with sparse singular vectors, resulting in a matrix which mostly consists of zeros.}
	\label{fig:sparsesvs}
\end{figure}

In particular, the most immediate challenge we encounter when developing a theory of matrix completion is that it is no longer possible to obtain the kind of uniform guarantees that apply to {\em all} low-rank matrices described in Section~\ref{sec:gauss}.  To see why, consider a matrix with rank one, but where one (or both) singular vectors are {\em sparse}, meaning that their energy is concentrated on just a few entries.  When this occurs, as illustrated in Figure~\ref{fig:sparsesvs}, the resulting matrix will also have its energy mostly concentrated on just a few entries, in which case most entries give us very little information and the recovery problem is highly ill-posed unless nearly all the entries are observed.  (Another way to see this is to realize that if only a few entries of such a matrix are observed, the matrix is very likely to live in the nullspace of the measurement operator $\cA$.)   More generally, if any particular column (or row) is approximately orthogonal to the span of the remaining columns (or rows), then it will be impossible to estimate without essentially observing the entire column (or row).

Thus, there exist low-rank matrices which, in the absence of any prior information, would clearly be impossible to recover from the observation of only a few entries.  In order to avoid such cases, we need to ensure that each entry of the matrix tells us something about the other rows/columns.  Fortunately, this is very likely to be the case in most practical applications of interest.  For example, the entire premise of the collaborative filtering approach to recommendation systems is that knowing how a particular user feels about a particular item provides information about other users and items.  This notion can be mathematically quantified in various different ways, but the bulk of the literature on matrix completion builds on the notion of {\em coherence} as introduced in~\cite{CandeR_Exact}.  Given an $R$-dimensional subspace $U$ of $\real^N$, the coherence is defined as
\begin{equation} \label{eq:coherence}
\mu(U) := \frac{N}{R} \max_{1 \le i \le N} \| \cP_U \ve_i \|_2^2,
\end{equation}
where $\{\ve_i\}$ denotes the canonical basis for $\real^N$ and $\cP_U$ denotes the orthogonal projection onto $U$. Clearly, the largest value that $\mu(U)$ can take is $N/R$, which occurs when some $\ve_i$ lies in the span of $U$.  The smallest possible value for $\mu(U)$ is 1, which occurs when $U$ is the span of vectors with constant magnitude $1/\sqrt{N}$.   %{\color{red} {\bf Figure?}}
Matrices whose column and row spaces have a small coherence (close to 1) are called {\em incoherent} and represent a class of matrices where each entry contains a comparable amount of information so that completion from a small number of observations is potentially feasible.

\subsection{Recovery guarantees}
There is now a rich literature providing a range of guarantees under which it is possible to recover a matrix $\mX_0$ from randomly chosen entries under the assumption that $\mX_0$ is incoherent and/or satisfies certain similar conditions. As a representative example, we will describe the guarantees that are possible when using nuclear norm minimization as a recovery technique, as first developed in~\cite{CandeR_Exact} and further refined in~\cite{Gross_Recovering,CandeT_Power,Recht_Simpler}.

To state the main conclusion of this literature, we will assume for the moment that $\mX_0$ is an $M \times N$ matrix of rank $R$ with singular value decomposition $\mU \mSigma \mV^*$. We will also assume that $\mu(\mU), \mu(\mV) \le \mu_0$ and further that the matrix $\mE = \mU \mV^*$ has a maximum entry bounded by $\mu_1 \sqrt{R/MN}$ .\footnote{As argued in~\cite{CandeR_Exact}, as a simple consequence of the Cauchy-Schwarz inequality, the assumption on $\mE$ will always hold with $\mu_1 = \mu_0 \sqrt{R}.$ In addition, a more refined analysis using alternative concentration bounds can eliminate the need for any assumption on $\mu_1$ (see, for example, \cite{Chen_Incoherence}). Moreover, given a limited amount of {\em a priori} information about the underlying matrix, it is also possible to obtain results that omit any dependence on the incoherence by sampling certain rows/columns more heavily~\cite{ChenBSW_Completing}. } We will also assume, without loss of generality, that $M>N$.  Then if $L$ entries of $\mX_0$ are observed with locations sampled uniformly at random with
\begin{equation}
\label{eq:MCmeasbound}
L \gtrsim \max( \mu_0, \mu_1^2) R(M+N) \log^2( M ),
\end{equation}
$\mX_0$ will be the solution to~\eqref{eq:nneq} with high probability.  We note that the required number of observations in the matrix completion setting exceeds that required in the Gaussian case (i.e., $O(R(M+N))$) in two natural ways.  First,  $L$ scales with the level of incoherence as quantified by $\mu_0$ and $\mu_1$.  For example, in the case where $\mu(\mU)$ or $\mu(\mV)$ approach their maximal value then the bound in~\eqref{eq:MCmeasbound} reduces to the requirement that we observe nearly every entry.  The second difference is that we have an additional $\log$ factor.  While the power of $2$ on the $\log$ may not be strictly necessary, some logarithmic dependence on the dimension of the matrix is a necessary consequence of the random observation model.  In particular, as a consequence of the classic {\em coupon collector} problem, we need at least $O(M \log M)$ observations simply to ensure that we observe each row at least once.  (If $N>M$, the same argument applies with columns instead of rows.)

Finally, we also note that the above result applies to the specific case of noise-free observations of a matrix with rank at most $R$.  Similar results can also be established that guarantee approximate recovery in the case of noisy observations and approximately low-rank matrices with the amount of recovery error being naturally determined by the amount of noise and degree of approximation error~\cite{KeshaMO_MatrixNoise, NegahW_Restricted, KoltcLT_Nuclear, CandeP_Matrix, RohdeT_Estimation, Klopp_Rank, GaiffL_Sharp, Klopp_High, Koltc_Von}.

The full proof of the exact recovery result described above is somewhat involved, but has been significantly simplified in the work of~\cite{Gross_Recovering} and subsequently in~\cite{Recht_Simpler} to the point where the key ingredients are relatively straightforward.  In light of the discussion of Section~\ref{ssec:MRIP} one might expect that a possible avenue of attack would be to show that by selecting elements of our matrix at random, we obtain a measurement operator $\cA$ that satisfies something like the matrix-RIP in~\eqref{eq:matrix-rip} but which holds only for matrices satisfying our incoherence assumptions.  In fact, it is indeed possible to pursue this route (and the incoherence assumption is vital to ensure that $\E[\|\cA(\mX)\|_2^2] = \|\mX\|_F^2$ as is required at the outset in Section~\ref{sec:gauss}).  This is essentially the approach taken in~\cite{NegahW_Restricted}.  However, the difference between two incoherent matrices is not necessarily itself incoherent, which leads to some significant challenges in an RIP-based analysis.  Moreover, this approach fails to yield the kind of exact recovery guarantee described above in the exactly low rank case and is primarily of interest in the noisy setting.

Instead of approaching the problem from the perspective of the matrix-RIP, the approach taken in~\cite{Gross_Recovering,CandeR_Exact, CandeT_Power,Recht_Simpler} relies on an alternative approach based on duality theory.  At a high level, the argument consists primarily of two main steps.  In the first step, one shows that if a dual vector satisfying certain special properties\footnote{Specifically, the required properties can be defined as follows. Given the matrix $\mX_0 = \mU \mSigma \mV^*$, define $T$ to be the subspace of $\real^{M \times N}$ spanned by elements of the form $\mU \mY^*$ and $\mX \mV^*$, where $\mX$ and $\mY$ are arbitrary, and let $\cP_T$ and $\cP_{T^\perp}$ denote the projections onto $T$ and its orthogonal complement, respectively.  Then our goal is to show that there exists a $\vlambda$ in the range of $\cA$ satisfying $\cP_T(\cA^*(\vlambda)) = \mU \mV^*$ and $\| \cP_{T^\perp}(\cA^*(\vlambda)) \| < 1$.} exists, then this is a sufficient condition to ensure exact recovery, i.e., that the minimizer of~\eqref{eq:nneq} is given by $\mX_0$.  This vector is called a {\em dual certificate} because it certifies that the (unique) optimal solution to~\eqref{eq:nneq} is $\mX_0$. This step in the argument relies entirely on elementary inequalities and standard notions from duality theory.

The more challenging part of the argument is the second step, in which one must show that given random samples of a matrix $\mX_0$ satisfying the required coherence properties, such a dual certificate must exist (with high probability).  The original proof of this in~\cite{CandeR_Exact} and the subsequent improvement in~\cite{CandeT_Power} involved rather intricate analysis, made especially difficult by the fact that when we observe $L$ distinct entries of a matrix, our observations are not fully independent (since we cannot sample the same entry twice).  In~\cite{Gross_Recovering} this analysis is dramatically simplified by two observations: {\em (i)} it is actually sufficient to merely obtain an {\em approximate} dual certificate, which can be easier to construct, and {\em (ii)} one can alternatively consider an observation model of sampling {\em with replacement}.  Under this model, one can analyze the adjoint operator $\cA^*(\cdot)$ by treating its output as a sum of {\em independent} random matrices, which enables the use of the powerful concentration inequalities recently developed in~\cite{AhlswW_Strong} that bound the deviation of this sum from its expected value. This allows one to analyze the relatively simple ``golfing scheme'' of~\cite{Gross_Recovering}, which consists of an iterative construction of an approximate dual certificate.  See~\cite{Recht_Simpler} for a condensed description of this approach and~\cite{Tropp_User} for an overview (and other applications) of the matrix concentration inequalities used in this analysis. It is also worth noting that this line of analysis also provides theory for alternative low-rank matrix recovery scenarios beyond simply the matrix completion case.  Indeed, the arguments in~\cite{Gross_Recovering} were originally developed to address the problem of quantum state tomography, where the goal is to efficiently determine the quantum state of a system via a small number of observations~\cite{GrossLFBE_Quantum}.  This problem can be posed as a matrix recovery problem, but where the $\mA_i$ are constructed from Pauli matrices.  Perhaps somewhat surprisingly, the theory described above can also be readily adapted to this scenario.

Finally, we also note that similar guarantees hold for a range of alternative algorithmic approaches. For example, both the approaches based on alternating minimization~\cite{jain13lo,Hardt_Understanding,HardtW_Fast} and spectral methods with iterations similar to the proximal method~\cite{KeshaMO_Matrix} described in Section~\ref{sec:mralgs} have been shown to provide exact reconstruction under similar coherence assumptions, but at the cost of a slight increase in the required number of observations.  In particular, the results of~\cite{KeshaMO_Matrix,jain13lo,Hardt_Understanding} all involve a dependence on the condition number of $\target$ in which the number of required observations grows when the $R^{\text{th}}$ singular value gets too small.  By a clever iterative scheme,~\cite{HardtW_Fast} shows that it is possible to eliminate this dependence in the context of alternating minimization.  However, just as in the Gaussian measurement case, the best-known guarantees for alternating minimization also involve a polynomial dependence on the rank $R$ as opposed to the linear dependence which can be obtained using other approaches.  Nevertheless, the substantial computational advantages of these approaches means that they provide an attractive alternative in practice.

%----------------------------------------------------------------------------------
\section{Nonlinear Observation Models}
\label{sec:nonlinear}

Although the theoretical results described in Sections~\ref{sec:gauss} and~\ref{sec:mc} are quite impressive, there is an important gap between the observation model described by~\eqref{def:A} and many common applications of low-rank matrix recovery.  As an example, consider the matrix completion problem in the context of a recommendation system where $\mX_0$ represents a matrix whose entries each represent a rating for a particular user on a particular item.  In most practical recommendation systems (or indeed, any system soliciting any kind of feedback from people), the observations are ``quantized'', for example, to the set of integers between 1 and 5.  If we believe that it is possible for a user's true rating to be, for example, 4.5, then we must account for the impact of this ``quantization noise'' on our recovery.  Of course, one could potentially treat quantization simply as a form of bounded noise and rely on the existing stability guarantees mentioned in Sections~\ref{sec:gauss} and~\ref{sec:mc}, but this is somewhat unsatisfying because the ratings are not simply quantized --- there are also hard limits placed on the minimum and maximum allowable ratings.  (Why should we suppose that an item given a rating of 5 could not have a true underlying rating of 6 or 7 or 10? And note that if this is indeed the case, then the ``quantization error'' can be potentially extremely large.)  In such a situation, it can be much more advantageous to directly consider a nonlinear observation model of the form
\begin{equation}\label{def:Q}
y_i = Q \left( \langle \target, \bA_i \rangle + z_i \right),
\end{equation}
where $Q(\cdot)$ is a scalar function that captures the impact of quantization, or any other potential nonlinearity of interest.  We describe some concrete examples below.

\subsection{One-bit observations}

The inadequacy of standard low-rank matrix recovery techniques in dealing with this effect is particularly pronounced when we consider problems where each observation is quantized to a single-bit.  In particular, suppose that our observations are given by
\begin{equation}\label{def:Q1bit}
y_i = \begin{cases} +1 & \text{if~} \langle \target, \bA_i \rangle + z_i \ge 0 \\
-1 & \text{if~} \langle \target, \bA_i \rangle + z_i < 0. \end{cases}
\end{equation}
In such a case, the assumptions made in the standard theory of matrix recovery do not apply, standard algorithms are ill-posed, and an alternative theory is required.  To see why, simply observe that in the noise-free setting (where $z_i = 0$), we could rescale $\target$ arbitrarily without changing any observations.\footnote{In fact, in the noise-free setting the situation is even worse than one might suspect. Even if the normalization is fixed/known a priori, the problem remains highly ill-posed.  See~\cite{DavenPBW_One} for further discussion.}

What is perhaps somewhat surprising is that when considering the noisy setting the situation completely changes --- the noise has a ``dithering'' effect and the problem becomes well-posed.  In fact, it is possible to show that one can sometimes recover $\target$ to the same degree of accuracy that is possible when given access to completely unquantized measurements, and that given sufficiently many measurements it is possible to recover $\target$ to an arbitrary level of precision. For the specific case of matrix completion, this observation model is analyzed in~\cite{DavenPBW_One} with the main conclusion being that it is possible to recover any $\target$ belonging to a certain class of approximately low-rank matrices up to an error proportional to $\sqrt{R(M+N)/L}$, so that by taking $L = C R(M+N)$ one can drive the recovery error to be arbitrarily small. The recovery algorithm in~\cite{DavenPBW_One} is a simple modification of~\eqref{eq:matrixlasso}, but where the fidelity constraint $\|\cA(\mX)-\vy\|_F^2$ is replaced by the negative log-likelihood of $\mX$ given observations from the model in~\eqref{def:Q1bit}.  A limitation of this approach is that it requires knowledge of the distribution of the noise $z_i$, but note that for many common noise distributions this still results in a convex optimization problem which can be solved with variants of the same algorithms described in Section~\ref{sec:mralgs}.  Similar results are described in~\cite{CaiZ_Max}, which again uses a penalized maximum-likelihood estimator but with a different regularizer than the nuclear norm.  See also~\cite{BhaskJ_1bit} and~\cite{SoniJHG_Noisy} which suggest potential improvements for the case of exactly low-rank matrices.  Note that many of these results are of interest even in the case where every entry of the matrix is observed, as this provides a theoretically justified way to reveal the ``underlying'' low-rank matrix given only quantized data.

While most of the existing research into the one-bit observation model has focused on the model in~\eqref{def:Q1bit}, and specifically in the matrix completion setting, it is important to note that it would not be difficult to extend this literature to handle the Gaussian observation model described in Section~\ref{sec:gauss} and/or related one-bit observation models.  As an example, an alternative model might involve a setting where we can only tell if the magnitude of our observations is ``small'' or ``large'', as captured by a model along the lines of
\begin{equation}\label{def:Qnearfar}
y_i = \begin{cases} +1 & \text{if~} |\langle \target, \bA_i \rangle + z_i| \ge T \\
-1 & \text{if~} |\langle \target, \bA_i \rangle + z_i| < T. \end{cases}
\end{equation}
A similar theory could be developed for this setting.  See~\cite{KarbaO_Robust} for example applications and one possible approach.

\subsection{Comparisons}

Another application of the one-bit observation models in~\eqref{def:Q1bit} or~\eqref{def:Qnearfar} arises when one considers scenarios involving comparisons between different entries in the matrix $\target$.   This might occur in the context of paired comparisons, where a person evaluates a pair of items and indicates whether they are similar or dissimilar, or whether one
is preferred to the other. This type of data frequently arises when dealing with judgements made by human
subjects, since people are typically more accurate and find it easier to make such judgements than to assign
numerical scores~\cite{David_Method}.  These settings can be readily accommodated by the models in~\eqref{def:Q1bit} or~\eqref{def:Qnearfar} by slightly modifying the $\mA_i$.  In particular, when comparing $X_0[m,n]$ to $X_0[m',n']$ one can set $\mA_i = \ve_m \ve_n^T - \ve_{m'} \ve_{n'}^T$ so that, for example,~\eqref{def:Q1bit} becomes
\begin{equation}\label{def:Qcompare}
y_i = \begin{cases} +1 & \text{if~} X_0[m,n]  + z_i \ge X_0[m',n'] \\
-1 & \text{if~} X_0[m,n]  + z_i < X_0[m',n']. \end{cases}
\end{equation}
A similar theory can be developed for low-rank recovery under this model.  For example, see~\cite{LuN_Individualized,ParkNZSD_Preference} for discussion of paired comparisons and~\cite{OhTX_Collaboratively} for a generalization to ordinal comparisons among groups of entries in $\target$.

\subsection{Categorical observations and other noise models}

While the discussion above has focused on the one-bit case where $Q(\cdot) = \text{sign}(\cdot)$, the algorithms developed for this case can often be readily extended to handle more general nonlinearities.  For example, in the case of more general quantization schemes to arbitrary finite alphabets, one must simply be able to compute the log-likelihood function in order to be able to compute the regularized maximum-likelihood estimate as analyzed in~\cite{DavenPBW_One}.  This has applications to multi-bit quantization schemes (e.g., ratings being quantized to the integers from 1 to 5) and also to handle general forms of categorical data (e.g., group membership, race/ethnicity, zip code, etc.).   Moreover, in many cases the analysis can also be extended to handle this kind of observation~\cite{LafonKMS_Probabilistic,KloppLMS_Adaptive,CaoX_Categorical}.  Finally, it is also worth noting that this body of work has also established a variety of both algorithmic and analytical techniques for handling a variety of complex probabilistic observation models.  As a result, this has laid the foundation for a number of works which consider more general noise models beyond simple bounded perturbations or Gaussian measurement noise.  For example,~\cite{CaoX_Poisson} explores the impact of the signal-dependent (Poisson) noise  that arises in dealing with applications involving count data.  These techniques are further generalized in~\cite{GunasRG_Exponential,Lafon_Low} to general exponential noise families. See also~\cite{KabasKMSZ_Phase} for a treatment of matrix factorization problems allowing for a quite general family of noise models. 

%----------------------------------------------------------------------------------
\section{Lifting}
\label{sec:lifting}

The progress in recovering a low-rank matrix from an incomplete set of linear measurements described above has also affected the way we think about solving {\em quadratic} and {\em bilinear} systems of equations.  There is a simple, but perhaps until recently under-appreciated, way to re-cast a system of quadratic equations as a system of linear equations whose solution obeys a rank constraint.  This method, known as {\em lifting}, is best illustrated with a small, concrete example.  Consider the following system of $6$ quadratic equations in three unknowns $v_1,v_2,v_3$:
\begin{equation*}
    \twoCol{
    \begin{aligned}
		4v_1^2 + 12 v_2^2 + 7v_3^2 + 25v_1v_2 + 16v_1v_3  + 7 v_2v_3  &= 237 \\
		13v_1^2 - 3v_2^2 + 2v_3^2 - 5v_1v_2 + 23v_1v_3 + 3v_2v_3 &= -4 \\
		-3v_1^2 - 12v_2^2 + 2v_3^2 -37v_1v_2 + 12v_1v_3 -6v_2v_3 &= -100 \\
		6v_1^2 + v_2^2 + 2 v_3^2 - 16 v_1v_2 + v_1v_3  + 3v_2v_3  &= 153 \\
		5v_1^2 - 25v_2^2 - 7v_3^2 + 8v_1v_2 - 4v_1v_3 - 4 v_2v_3 &= -459 \\
		-9v_1^2 -9v_2^2 + 4v_3^2 -20v_1v_2 -2v_1v_3 + 10v_2v_3 &= 230.
	\end{aligned}}
    {\begin{aligned}
		4v_1^2 + 12 v_2^2 + 7v_3^2 + 25v_1v_2 + 16v_1v_3  + 7 v_2v_3  &= 237 \\
		13v_1^2 - 3v_2^2 + 2v_3^2 - 5v_1v_2 + 23v_1v_3 + 3v_2v_3 &= -4 \\
		-3v_1^2 - 12v_2^2 + 2v_3^2 -37v_1v_2 + 12v_1v_3 -6v_2v_3 &= -100
	\end{aligned}
	\quad
	\begin{aligned}
		6v_1^2 + v_2^2 + 2 v_3^2 - 16 v_1v_2 + v_1v_3  + 3v_2v_3  &= 153 \\
		5v_1^2 - 25v_2^2 - 7v_3^2 + 8v_1v_2 - 4v_1v_3 - 4 v_2v_3 &= -459 \\
		-9v_1^2 -9v_2^2 + 4v_3^2 -20v_1v_2 -2v_1v_3 + 10v_2v_3 &= 230.
	\end{aligned}}
\end{equation*}
These equations are quadratic in the entries of the vector $\vv$, but they are linear in the entries of the matrix
\[
	\mX = \vv\vv^\T =
	\begin{bmatrix}
		v_1^2 & v_1v_2 & v_1v_3 \\
		v_2v_1 & v_2^2 & v_2v_3 \\
		v_3v_1 & v_3v_2 & v_3^2
	\end{bmatrix}.
\]
For example, the first two equations above can be written as
\twoCol{
\[
    \inner{\mX,\mA_1}=237 \quad\text{and}\quad \inner{\mX,\mA_2}=-4,
\]
with
\[
	\mA_1 =
	\begin{bmatrix}
		4 & 12.5 & 8 \\ 12.5 & 12 & 3.5 \\ 8 & 3.5 & 7
	\end{bmatrix}, ~~
	\mA_2 =
	\begin{bmatrix}
		13 & -2.5 & 11.5 \\ -2.5 & -3 & 1.5 \\ 11.5 & 1.5 & 2
	\end{bmatrix}.
\]
}{
\[
    \inner{\mX,\mA_1}=237,\quad\text{with}\quad
	\mA_1 =
	\begin{bmatrix}
		4 & 12.5 & 8 \\ 12.5 & 12 & 3.5 \\ 8 & 3.5 & 7
	\end{bmatrix},
	\quad
	\inner{\mX,\mA_2}=-4,\quad\text{with}\quad
	\mA_2 =
	\begin{bmatrix}
		13 & -2.5 & 11.5 \\ -2.5 & -3 & 1.5 \\ 11.5 & 1.5 & 2
	\end{bmatrix}.
\]}
The other four measurement matrices $\mA_3,\ldots,\mA_6$ are defined similarly.  Using these six matrices to define $\cA:\real^{3\times 3}\rightarrow\real^6$ along with $y_1=-237,\ldots,y_6 = 230$, we compute
\[
	\minimize_{\mX} \|\mX\|_*\quad\text{subject to}\quad \inner{\mX,\mA_\ell}=y_\ell,~~\ell=1,\ldots,6,
\]
which has solution
\[
	\mX^\star =
	\begin{bmatrix}
		1 & -3 & 5 \\ -3 & 9 & 15 \\ -5 & 15 & 25
	\end{bmatrix}
	=
	\begin{bmatrix}
		-1 \\ 3 \\ 5
	\end{bmatrix}	
	\begin{array}{@{}c@{}}{
  		\begin{bmatrix}
    			-1 & 3 & 5
  		\end{bmatrix}}\\\\
	\end{array}.
\]
The reader can verify that $v_1=-1,v_2=3,v_3=5$ is indeed a solution to the equations above.  This solution is also unique up to a sign change.

% bilinear ?

We see now that solving any system of quadratic equations is equivalent to finding a rank-1 matrix that obeys a set of linear equality constraints.  The low-rank recovery results discussed above can now be applied directly to the problem of solving such equations.  For example, we have seen in Section~\ref{sec:gauss} that a linear operator defined by $\N\times\N$ matrices $\mA_1,\ldots\mA_{\obs}$ whose entries are populated by independent random variables will stably embed the set of rank-1 matrices when $\obs\geq\mathrm{Const}\cdot\N$.  This means that if we have a slightly over-determined system of quadratic equations with random coefficients which does in fact have a solution, then we can find that solution by applying any one of a number of low-rank recovery algorithms.

%-----------------------------------------------------------------------------
\subsection{Phase retrieval}

Quadratic and bilinear problems that arise in applications typically have special structure in the equation coefficients $\mA_m$.  One type of problem that has received considerable attention in the literature recently is recovering a vector $\vv_0\in\R^{\N}$ from observations of the {\em magnitude} of a series of linear measurements:
\begin{equation}
	\label{eq:magmeas}
	y_\ell = |\inner{\vv_0,\va_\ell}|,\quad \ell=1,\ldots,\obs.
\end{equation}
If the $\{\va_\ell\}$ span $\R^\N$ (which will happen for a generic set of measurement vector when $\obs\geq\N$), then knowing the signs (or phase, if the vectors are complex) of the
$\inner{\vv_0,\va_\ell}$ along with the $y_\ell$ would be enough to recover $\vv_0$ --- for this reason, this problem is referred to as {\em phase retrieval} in the literature.

By squaring the observations in \eqref{eq:magmeas}, we see that
\begin{equation}
	\label{eq:lifting}
    \twoCol{
    |\inner{\vv_0,\va_\ell}|^2 = \vv_0^\T\va_\ell\va_\ell^\T\vv_0 = \inner{\mX_0,\mA_\ell},}
    {|\inner{\vv_0,\va_\ell}|^2 = \vv_0^\T\va_\ell\va_\ell^\T\vv_0 = \inner{\vv_0\vv_0^\T,\va_\ell\va_\ell^\T} = \inner{\mX_0,\mA_\ell},}
\end{equation}
where both $\mX_0=\vv_0\vv_0^\T$ and $\mA_\ell =\va_\ell\va_\ell^\T$ are rank 1 and symmetric.  This means that the equation coefficients in the quadratic phase retrieval problem are structured in that they can be arranged into rank 1 matrices.

In \cite{balan06si}, it is shown that for generic $\va_\ell$, if $\obs\geq 2N-1$, then $\vv_0$ is the only vector (up to a sign) that has the measurements \eqref{eq:magmeas}.  This bound is tight, as for $\obs < 2N-1$ there is no set of $\va_\ell$ such that $\vv_0$ is uniquely specified.\footnote{The threshold becomes $\obs\lessgtr 4N-2$ when the vectors are complex.}
The connection between phase retrieval and low-rank recovery was popularized in~\cite{candes13ph}, which gave the name {\em PhaseLift} to the method of treating the nonlinear inverse problem \eqref{eq:magmeas} as the rank-constrained linear inverse problem \eqref{eq:lifting} and showed that
\begin{align*}
	\minimize_{\mX}~\|\mX\|_* ~~ \text{subject to} ~~ \inner{\mX,\mA_\ell}&=y_\ell^2,~\ell=1,\ldots,\obs, \\
	\mX&\succeq\mzero,
\end{align*}
successfully recovers $\mX_0=\vv_0\vv_0^\T$ when $\obs\gtrsim\N\log\N$.  In \cite{demanet14st}, it was shown that the nuclear norm minimization above is actually unnecessary: the rank-1 $\mX_0$ is the only matrix that is both in the SDP cone ($\mX_0\succeq\mzero$) and obeys the linear constraints.  The recovery task, then, is just to find a matrix that obeys these sets of constraints --- \cite{demanet14st} also shows that a straightforward projection onto convex sets (POCS) algorithm has linear convergence.  Recent work in \cite{candes14so} has refined these results to show that $\obs\gtrsim N$ observations are sufficient.

%---
Once the $\va_\ell$ are chosen, they will work {\em uniformly} well (with high probability) for all vectors $\vv_0$ \cite{kueng14lo}.  That work also shows that the uniform recovery results extend to target matrices $\mX_0$ with general rank $\R$.  A related result appears in \cite{chen15ex}, which examines this problem in the context of estimating a (low-rank) covariance matrix from {\em sketches} of the data vectors.

As with many of the analytical results we have discussed, the randomness of the $\{\va_\ell\}$ plays a large role in establishing the effectiveness of the recovery technique.  In practice, the types of measurements that can be made are determined by the physics of the measurement system, and taking the $\va_\ell$ to be isotropic random vectors is not a realistic model.  However, it is sometimes possible to alter the acquisition system slightly to inject enough diversity in the $\va_\ell$ to make the recovery effective.  This is analyzed in detail in \cite{candes15ph}, where the acquisition process consists of first modulating a signal with a small number of pseudo-random patterns, then measuring the magnitudes of the Fourier transform of the result.  Another example can be found in \cite{eldar15sp}, where magnitudes of the short-time Fourier transform are measured, a problem which arises in several different applications in signal processing and optics.

The lifting framework described above is not the only path to a convex relaxation.  The {\em PhaseCut} method, proposed in \cite{waldspurger15ph},  casts the recovery problem as an equality constrained quadratic program, and then uses a well-known  relaxation for this type of problem.  Interestingly, although the PhaseCut and PhaseLift relaxations are in general different, they recover the correct $\mX_0=\vv_0\vv_0^\T$ under identical conditions. %\cite{???}

Finally, there are mathematical guarantees for other algorithms for solving the nonlinear inverse problem in \eqref{eq:magmeas}.  Algorithms very similar to the alternating minimization algorithm from Section~\ref{sec:alternatingprojections} have been used for phase retrieval since the 1970s \cite{fienup82ph}.  In \cite{netrapalli13ph}, a theoretical analysis of alternating projections for phase retrieval was given (along with an intelligent method for initializing the algorithm) that gives a guarantee of effectiveness when $\obs\gtrsim \N\log^3\N$.  Recent work in \cite{candes14ph} shows that a local descent algorithm again coupled with a smart initialization is as effective for phase retrieval as the convex relaxations above while being far more computationally efficient.
Algorithms of this type also have performance guarantees for recovering rank-$R$ matrices from measurements of the form \eqref{eq:lifting} \cite{WhiteSW_Local}.

An extended survey of the recent work on this problem, including a more in depth discussion of most of the results above, can be found in \cite{shechtman15ph}.

%-----------------------------------------------------------------------------
\subsection{Blind deconvolution}

\begin{figure}
	\centering
	\includegraphics[scale=0.25]{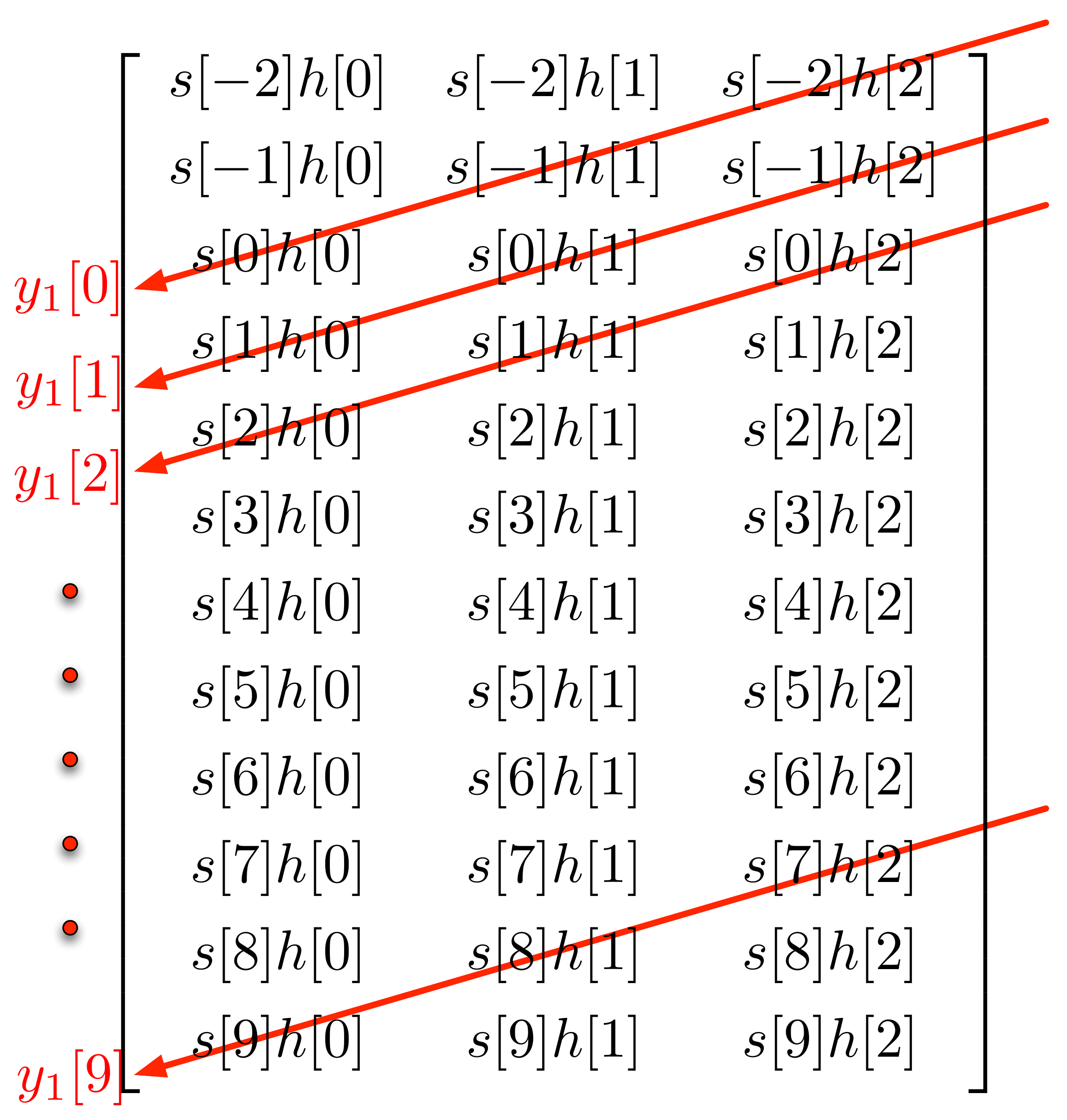}
	\caption{\small\sl Each sample of the convolution $\vy=\vs\star\vh$ in \eqref{eq:convolution} is a sum along a skew-diagonal of the rank-1 matrix $\vs\vh^\T$.}
	\label{fig:matrixtomo1}
\end{figure}

Perhaps an even more important and prevalent problem, especially in the signal processing and communications communities, is {\em blind deconvolution}.  Here we observe the (discrete-time) convolution of two signals,
\begin{equation}
	\label{eq:convolution}
	y[\ell] = \sum_{n}h[n]s[\ell-n],
\end{equation}
and want to recover $\vh$ and $\vs$.  Each sample of $\vy$ is a different linear combination of entries in $\vs$ multiplied by entries in $\vh$, and so this is a {\em bilinear} inverse problem.  It can be recast (``lifted'') into a rank-constrained linear inverse problem in almost exactly the same way described above for quadratic problems.  The only difference is that we want to recover a non-symmetric rank-1 matrix.  For the convolution in \eqref{eq:convolution}, each sample is the sum along a skew-diagonal of $\vs\vh^\T$; this is illustrated in Figure~\ref{fig:matrixtomo1}.

%----
%----

This is in general a intractable problem without at least partial information about $\vh$ and/or $\vs$.  We will start with only the simplest of assumptions --- that $h[k]$ takes non-zero values only on $k=0,\ldots,K-1$ and that $s[n]$ takes non-zero values only on $n=0,\ldots,\N-1$.  The convolution, then, will be non-zero only on $\ell = 0,\ldots,\obs-1=\N+K-1$.  In this case, if we translate our observations into the Fourier domain, then we can naturally express them as inner products against known rank-1 matrices.  The discrete-time Fourier transform (DTFT) of signal $x[n]$ with arbitrary length is given by
\[
	\hat{x}(\omega) = \sum_{n=-\infty}^\infty x[n]\e^{-\j\omega n},\quad -\pi\leq\omega<\pi.
\]	
The DTFT of the observations $\vy$, after we zero-pad it outside its support, is the point-by-point multiplication of the DTFTs of $\vs$ and $\vh$ (also after zero-padding).  Since all three signals are zero outside of $\{0,\ldots,\obs-1\}$, we can evaluate the DTFT of the observations at $\obs$ equally spaced frequencies between $-\pi$ and $\pi$ by computing the vector inner products
\[
	\hat{y}[\ell] := \hat{y}(\omega_\ell) = \inner{\vy, \vf_{\ell,\obs}}, 
\]
where $\omega_\ell = 2\pi(\ell-1)/\obs$ for $\ell=0,\ldots,\obs-1$ and $\vf_{\ell,\obs}$ is the Fourier vector in $\complex^\obs$ that has entries $f_{\ell,\obs}[n] = \e^{\j\omega_\ell n},~n=0,\ldots,\obs-1$.  All $\obs$ of these inner products can be produced quickly by applying a fast Fourier transform (FFT) to the vector $\vy$.  Using $\vf_{\ell,\N}\in\complex^\N$ and $\vf_{\ell,K}\in\complex^K$ for length $\N$ and $K$ Fourier vectors (still at frequencies $\omega_\ell$), we can write
\begin{align*}
	\hat{y}[\ell] &= \hat{s}(\omega_\ell)\cdot \hat{h}(\omega_\ell) \\
	&= \left(\sum_{n=0}^{N-1}s[n]\e^{-\j\omega_\ell n}\right)\left(\sum_{k=0}^{K-1}h[k]\e^{-\j\omega_\ell k}\right) \\
	&= \inner{\vs,\vf_{\ell,N}}\inner{\vh,\vf_{\ell,K}} \\
	&= \inner{\vs\vh^\T,\vf_{\ell,N}\conj{\vf}_{\ell,K}^\H}.
\end{align*}
For real-valued $\vs$ and $\vh$, the real and imaginary parts of $\hat{y}(\omega_\ell)$ yield inner products against two real-valued rank-1 matrices:
\[
	\Re{\hat{y}[\ell]} = \inner{\vs\vh^\T,\mA_\ell},\quad \Im{\hat{y}[\ell]} = \inner{\vs\vh^\T,\mB_\ell},
\]
where $\mA_\ell,\mB_\ell$ are $\N\times K$ matrices with entries
\[
	A_\ell[n,k] = \cos(\omega_\ell(n+k)),
	\quad
	B_\ell[n,k] = -\sin(\omega_\ell(n+k)).
\]

With only the finite support assumption, the blind deconvolution problem is still fundamentally hard.  Past even the trivial ambiguity of replacing $\vs,\vh$ with $\alpha\vs,\alpha^{-1}\vh$ for some $\alpha\not=0$, there will typically be many pairs of vectors consistent with the measurements $\vy$ --- the argument for this is carefully laid out for linear convolution in the recent work \cite{choudhary14fu}.  However, more generic constraints on $\vs$ and $\vh$ can make the problem very well-posed, and allow it to be solved using convex relaxations.  In \cite{ahmed14bl}, it is shown that if $\vh$ is restricted to have any known support of size $K$ on $\{0,\ldots,\obs\}$, and $\vs$ comes from a known random subspace of dimension $\N$, then nuclear norm minimization will recover $\vs\vh^\T$ with high probability when
\[
	\max(K,\N) \leq \mathrm{Const}\cdot \frac{\obs}{\log^3\obs}.
\]
This says that the dimension of our linear models (which determines their expressive power) can be within a logarithmic factor of the ambient dimension.  General identifiablility results for deterministic subspace models are discussed in \cite{LiLB_IdentifiabilityA}.

% A similar result exists for the multichannel problem ...

The lifting technique described above for blind deconvolution is just one of many methods that have been proposed for this important problem (see the books \cite{haykin94bl,campisi07bl} or the survey \cite{tong98mu} for an overview of algorithms used in digital communications, imaging, and other areas).  The lifting technique, however, puts the problem into the realm of optimization.  This makes it very natural to add (convex) constraints for modeling prior knowledge about the signal, or integrate indirect or partial measurements.  In \cite{tang14co,bahmani15li}, for example, it is shown that if the image is modulated before being blurred by an unknown kernel, the recovery problem is actually very well-posed.  Recovery in this scenario is possible without any prior knowledge of the image, and the restrictions on the blur kernel are very mild.  In \cite{LingS_Selfcalibration}, the lifting framework is applied to the closely related problem of auto-calibrating sensor arrays.  Techniques for encouraging $\vs$ and/or $\vh$ to be sparse have recently been studied in \cite{LeeLJB_Stability,Chi_Guaranteed,bahmani15ne}.

%-----------------------------------------------------------------------------
% If we decide to do a section on Simulataneously Low Rank and Sparse, this can be the seed
%{\color{red} \subsection{Simultaneous structure (?)}
%
%Recover $\vs\vh^\T$ where one or both of $\vs,\vh$ is sparse ...\\[2mm]
%
%Obvious relaxation is suboptimal (Oymak et al) ... \\[2mm]
%
%Known results exploit sparsity or low-rank, not both (Li and Voroninski, Chi/Goldsmith) ...\\[2mm]
%
%There are nonconvex algorithms with provable guarantees (Li/Bresler sparse power factorization), but these require restricted isometries and so do not necessarily work with the applications mentioned above ...\\[2mm]
%
%If the measurement $\va_\ell$ are in a low-dimensional subspace, then a simply two-stage convex optimization procedure is near-optimal (Bahmani) ...\\[2mm]
%
%}

\frenchspacing
\bibliographystyle{plain}
\bibliography{MDbib,lrr-refs}

\end{document}